\documentclass[11pt,twoside,reqno]{amsart}
\usepackage{epsfig}
\calclayout 

\begin{document}
\title{Open-Loop Control of Quantum Particle Motion: Effective Splitting in Momentum Space}
\maketitle
\begin{center}
\author{Babar Ahmad$^{1}$, Sergei Borisenok$^{1,2}$, Saifullah$^{1}$, Yuri
Rozhdestvensky$^{3}$}\\
\vskip 0.2cm \small{{\enskip $^{1}$School of Mathematical Science,
GC University, Lahore, Pakistan\\ \enskip $^{2}$ Department
of Physics, Herzen State University, St. Petersburg, Russia\\
\enskip $^{3}$ Institute of Laser Physics, St. Petersburg, Russia}}
\end{center}
\begin{abstract}
In this paper an effective quantum particle beam-splitter in the
momentum space is realized in the frame of open-loop control scheme.
We demonstrate for small interaction time that the splitting effect
$\pm40\hbar k$ with summarized relative intensity in both main
components is about 50 per cent from initial intensity of the atomic
beam. \vskip 0.2 true cm
\noindent{\it Key words } : Open-loop control, beam splitter. \\
\noindent{\it AMS Subject Classification} : 35B30, 35B37.
\end{abstract}
\maketitle
\pagestyle{myheadings} \markboth{\centerline {\scriptsize Babar
Ahmad, Sergei Borisenok, Saifullah, Yuri Rozhdestvensky }}
{\centerline {\scriptsize  Open-loop Control of Quantum Particle
Motion:Effective Splitting in Momentum Space}}
\section{Introduction}
In nano-lithography the optical control of atomic motion is one of
the main problems. In principle, we can make such a control for atom
dynamics because there is an exchange of momentum between the atoms
and the optical fields. The momentum exchange can be created with
practical devices: atomic mirrors and atomic beam splitters which
are main elements of atomic interferometer. The most interesting
possibility here is to obtain the splitting of the initial atomic
wave packet coherently into two main momentum components only by
controllable way. It is needed both for increasing of atomic
interferometer sensitivity and for the creation of periodic
nano-structures by atomic wave packet lithography \cite{1}.
Previously, successful splitting of an atomic wave packet has been
achieved by using Raman pulses, magneto-optical beam splitter,
diffraction in an optical standing wave, adiabatic passage. More
recently, coherent splitting has been realized by scattering of an
atomic wave packet in standing wave with modulated intensity
\cite{2} and by using chirped standing wave fields \cite{3}.\\ In
this paper we concentrate on the possibility to split an atomic wave
packet in standing wave with modulated amplitude because this
beam-splitter has a number of advantages by comparing with others.
The first one is the simplicity for an experimental realization
because it is quite easy to obtain the time modulation of intensity
with any shape. The second advantage is that the scale splitting of
an atomic wave packet can be controlled by changing the values of
both an amplitude and the frequency of the modulation. Actually, we
demonstrate for small interaction time, which is required for clear
and large splitting, the beam-splitting effect $\pm40\hbar k$ with
summarized relative intensity in both main components about 50 per
cent from initial intensity of the atomic beam.\\
The principal opportunity to split the beam in the momentum space
was demonstrated in \cite{4}. To achieve the effective splitting, we
apply the scheme of open-loop control, or feed-forward control, i.e.
a control signal depends only on the time. Our control goal is to
obtain the large angle splitting for the initial wave packet after
some time of the interaction between the atoms and the field of
modulated standing wave.

\section{Physical background and mathematical model for beam splitting in the momentum space}
Atom lithography is an active field now a days. The resolution of an
optical lithography technology is limited by diffraction, which for
the case of deep ultraviolet light approaches 200 nm. The progress
of recent device technology requires smaller patterning of 10 nm
size. However, when one tries to make very small devices, the
resolution of the resistance is limited by the spread of the
secondary electron in an electron beam lithography, as well as in
X-ray lithography.\\
The ability to generate ultracold atoms using lasers has opened up
new possibilities. The long de Broglie wave of cold atoms makes
possible an interferometric manipulation with atomic wave packets,
which is designed by an optical standing wave. In this case, atoms
can be controlled directly to form a desired pattern. To produce the
pattern with high resolution, we need to split the wave packet into
two coherent momentum components only. For the model with only two
states (i.e. an approximation of two level atom states), we have to
split the population of the lower state (because the population of
the exited state usually loses the coherency very fast by
spontaneous decay) in several momentum components (in an ideal case
-- only two). At the same time to form the pattern with small step
we need to control the scale of splitting between two main coherent
components in momentum space. Therefore, an atomic beam splitter is
the main element for the practical realization of nano-scale
lithography with the controlled step by coherent scattering of an
atomic wave packet.\\
Let's consider now a two level atom in a far detuned standing wave
with the intensity modulated in time as $I=I_0f(\varepsilon ,\Delta
\cdot t, \phi _0) \mathrm{cos}(kx)$, where $\varepsilon $ is an
amplitude, $\Delta $ is the frequency of the modulation, and $\phi
_{0}$ is an initial modulation phase. The standing wave with the
frequency $\omega _{1}$ applies between two states of atom system,
where the state 1 is ground and the state 2 is the exited one. Here,
$\omega _{0}$ is the frequency of atom transition and the difference
$\omega _{1}-\omega _{0}$ is the detuning. We will assume that the
beam from an atom source propagates along $z$-axis and crosses the
optical wave, standing along $x$-axis, by right angle. The
spontaneous emission from the upper level in this system can not be
neglected. After some time $t$ of the interactions between the atoms
and the field of the standing wave, the initial atomic wave packet
is splitted in few coherent momentum components.\\
Dynamics of the atom in the modulated standing wave is described
with nonstationary Schr$\ddot{o}$dinger equation for the wave
function $\Psi (r,t)$ of the two level atom:
\begin{equation}
i\hbar \frac{\partial \Psi (r,t)}{\partial t}=\hat{H}\Psi (r,t),  \label{1}
\end{equation}
where ${\hat H}$ is a Hamiltonian which takes into account both the
atom movement along the standing wave and the dipole interaction
between the atom and the optical field. For sufficiently large
detuning, when it is much larger than Rabi frequency and the natural
width of the atomic transition, $\Omega >>R_{0},\Gamma $ (where
$R_{0}$ is the Rabi frequency, $\Gamma $ is the natural width of the
transition), the excited state 2 can be adiabatically eliminated. As
the result, we obtain the equation for the amplitude of the
probability of the ground state $\Psi _{1}(x,t)$:
\begin{equation}
i\hbar \frac{\partial \Psi _{1}(x,t)}{\partial t}=-\frac{\hbar
^{2}}{2m} \triangle _{xx}\Psi _{1}(x,t)+\frac{R_{0}^{2}}{\Omega
}[f(\varepsilon ,\Delta \cdot t, \phi
_0)]^{2}\mathrm{cos}^{2}(kx)\Psi _{1}(x,t),\   \label{2}
\end{equation}
and $m$ is the atom mass.\\
After the Fourier transform the same equation in the momentum space is given
by:
\begin{equation}
i\frac{\partial \Psi _{1}(p,\tau)}{\partial \tau}=(p^{2}+R^{2})\Psi
_{1}(p,\tau )+\frac{R^2}{2}\left[ \Psi _{1}(p+2,\tau )+\Psi
_{1}(p-2,\tau ) \right], \   \label{3}
\end{equation}
where
$
R^2 =\frac{R_0^2}{2\Omega \omega _R} \left[ f(\varepsilon, (\Delta /\omega
_R)\cdot\tau , \phi _0) \right] ^2 ,
$
$\omega _R = \hbar k^2/2m$ is called a recoil frequency, $\tau =
\omega _Rt.$ Here we normalized atom momentum along $x$-axis to
$\hbar k$ and the other quantities (the interaction time, the Rabi
frequency and the detuning) we normalized to the recoil frequency
$\omega _R$. We have to point out that equations (2)-(3) are valid
in the approximation when both the changing of atom momentum along
$z$-axis and the initial value along $x$-axis can be neglected.

\section{Shell model for the splitting process: parametric control}
To explain the effect of splitting in the momentum space we start
from the case of parametric control with a constant $R$. We invent a
complex shell model for $\Psi _{1}$-function. Initially the atomic
beam has a Gaussian distribution centered at $p=0$. Thus, from the
structure of the RHS (3) we can expect the non-zero meanings of
$\Psi _{1}$ functions to be concentrated in the neighborhoods of the
points $p=2n$, where $n=0,\pm 1,\pm 2,...$. Then we can predict the
continuous dependency on $p$ with the discrete number $n$: $\Psi
_{1}(p+2n,\tau )=y_{n}(\tau )$. Dynamics Eq. (3) can be re-written
in the form:
\begin{equation}
i\frac{dy_{n}(\tau )}{d\tau }=(4n^{2}+R^{2})y_{n}(\tau
)+\frac{R^{2}}{2}
\left[ y_{n-1}(\tau )+y_{n+1}(\tau )\right]
\label{4}
\end{equation}
with the initial conditions: $y_{0}(0)=1\; \ \ \ y_{n\not=0}(0)=0. $
Now we want to limit our shell number. In the case of three shells
only Eq. (4) becomes very simple:
\begin{gather}  \label{5}
i\frac{dy_0(\tau )}{d\tau}=R^2y_0(\tau) +R^2y_1(\tau ); \ \  \\
i\frac{dy_1(\tau )}{d\tau}=(R^2+4)y_1(\tau) +\frac{R^2}{2}y_0(\tau
). \ \  \notag
\end{gather}
We demand for the elder shells : $y_{\pm 2}=y_{\pm 4}=...\equiv 0$ for any
moment $\tau$. Eq. (5) can be easily solved:
\begin{eqnarray}  \label{6}
y_0(\tau )=\mathrm{e}^{-i(R^2+2)\tau}\left[ C_1 \mathrm{e}^{i\omega \tau}+
C_2\mathrm{e}^{-i\omega\tau}\right]; \ \  \\
y_1(\tau )=\frac{1}{R^2}\mathrm{e}^{-i(R^2+2)\tau}\left[ -(\omega
-2)C_1 \mathrm{e}^{i\omega \tau}+ (\omega
+2)C_2\mathrm{e}^{-i\omega\tau}\right], \ \   \notag
\end{eqnarray}
with $\omega = \sqrt{2R^4+16}/2$. For the initial conditions $y_0(0)=1$ and
$y_1(0)=0$ the constants are: $C_1=(\omega +2)/2\omega$ and $C_2=(\omega
-2)/2\omega$. The corresponding population amplitudes of the shells $0$ and
$\pm 1$ are given by:
\begin{eqnarray}  \label{7}
a_0(\tau )=y_0(\tau )y_0^*(\tau )=1
 - \frac{R^4}{R^4+8}\mathrm{sin}^2\left( \frac{\sqrt{2R^4+16}\tau}{2}
\right); \ \  \\
a_1(\tau )=y_1(\tau )y_1^*(\tau
)=\frac{R^4}{2R^4+16}\mathrm{sin}^2\left( \frac{\sqrt{2R^4+16}\tau
}{2}\right). \ \   \notag
\end{eqnarray}
Surely, the normalization $a_0+a_{-1}+a_{+1}=a_0+2a_1=1$ is saved
for any moment $\tau$. Now we can see that in the 3-shell model the
regular splitting is realized when the time $\tau = (2k+1)\pi
/\sqrt{2R^4+16}$, $k=0,1,2,........$ The same effect can be
reproduced in the case of five shells. To simplify the final
expression we will omit the coefficients 4 and 16 in RHS, because
the numerical meaning of $R^{2}$ is about 300 (i.e. $R^{2}>>4\
\mathrm{and}\ 16$). Then we apply Laplace transform:
\begin{eqnarray}
i(sY_{0}(s)-1) &=&R^{2}Y_{0}(s)+R^{2}Y_{1}(s)\; \   \notag  \label{8} \\
isY_{1}(s) &=&R^{2}Y_{1}(s)+\frac{R^{2}}{2}[Y_{0}(s)+Y_{2}(s)]\; \  \\
isY_{2}(s) &=&R^{2}Y_{2}(s)+\frac{R^{2}}{2}Y_{1}(s).\ \   \notag
\end{eqnarray}
Then:
\begin{eqnarray}
Y_{0}(s) &=&\frac{\left( s+\frac{3}{2}iR^{2}\right) \left( s+\frac{1}{2}%
iR^{2}\right) }{(s+iR^{2})\left( s+\frac{2-\sqrt{3}}{2}iR^{2}\right) \left(
s+\frac{2+\sqrt{3}}{2}iR^{2}\right) }\; \   \notag  \label{9} \\
Y_{1}(s) &=&-\frac{iR^{2}}{2\left( s+\frac{2-\sqrt{3}}{2}iR^{2}\right)
\left( s+\frac{2+\sqrt{3}}{2}iR^{2}\right) }\ ;\  \\
Y_{2}(s) &=&-\frac{R^{4}}{4(s+iR^{2})\left( s+\frac{2-\sqrt{3}}{2}%
iR^{2}\right) \left( s+\frac{2+\sqrt{3}}{2}iR^{2}\right).}\ \notag
\end{eqnarray}%
With the inverse Laplace transform we restore the time-dependent solution in
the momentum space:
\begin{eqnarray}
y_{0}(\tau ) &=&\frac{1}{3}\left[ \mathrm{exp}(-iR^{2}\tau )+\mathrm{exp}
\left( -\frac{iR^{2}\tau }{2(2+\sqrt{3})}\right) +\mathrm{exp}\left( -\frac{%
iR^{2}\tau }{2(2-\sqrt{3})}\right) \right] \; \   \notag  \label{10} \\
y_{1}(\tau ) &=&\frac{\sqrt{3}}{6}\left[ \mathrm{exp}\left( -\frac{%
iR^{2}\tau }{2(2-\sqrt{3})}\right) -\mathrm{exp}\left( -\frac{iR^{2}\tau }{%
2(2+\sqrt{3})}\right) \right] \; \  \\
y_{2}(\tau ) &=&-\frac{1}{3}\mathrm{exp}(-iR^{2}\tau )+\frac{1}{6}\left[
\mathrm{exp}\left( -\frac{iR^{2}\tau }{2(2+\sqrt{3})}\right) +\mathrm{exp}%
\left( -\frac{iR^{2}\tau }{2(2-\sqrt{3})}\right) \right] \ . \
\notag
\end{eqnarray}
The amplitudes are harmonical:
\begin{eqnarray}
a_{0}(\tau ) &=&y_{0}(\tau )y_{0}^{\ast }(\tau )=\frac{1}{9}\left[ 1+2%
\mathrm{cos}\left( \frac{\sqrt{3}}{2}R^{2}\tau \right) \right]
^{2}\; \
\notag  \label{11} \\
a_{1}(\tau ) &=&y_{1}(\tau )y_{1}^{\ast }(\tau )=\frac{1}{3}\mathrm{sin}%
^{2}\left( \frac{\sqrt{3}}{2}R^{2}\tau \right) \; \  \\
a_{2}(\tau ) &=&y_{2}(\tau )y_{2}^{\ast }(\tau )=\frac{1}{9}\left[ 1-\mathrm{%
cos}\left( \frac{\sqrt{3}}{2}R^{2}\tau \right) \right] ^{2}. \
\notag
\end{eqnarray}
Surely, again the normalization
$a_{0}+a_{-1}+a_{+1}+a_{-2}+a_{+2}=a_{0}+2a_{1}+2a_{2}=1$ is satisfied. The
splitting effect is obtained, when $\mathrm{cos}(\sqrt{3}R^{2}\tau /2)=-1$,
then the population of the $\pm 2$ shells is: $2a_{2}=8/9$, and in the same
time $a_{1}=0$ and $a_{0}=1/9$ only (see Fig.1).

\begin{figure}[h]
 \hskip0cm\centerline{\hbox{
 \includegraphics[clip=true, bb=0 1 15cm 11.1cm]{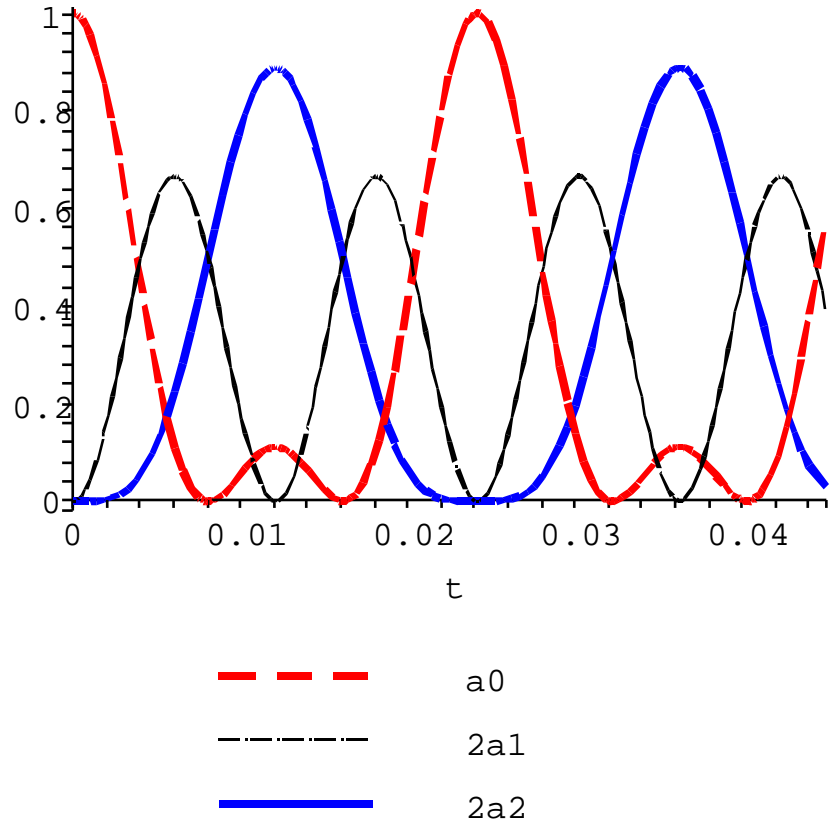}}}
  \caption{Splitting effect for the model of 5 shells.}
\label{fig1}
\end{figure}

However, if the number $n$ of a shell is increased such that
$4n^2>>R^2$ (i.e. for $R\simeq \sqrt{300}$ we have $n>>10$), then
$R^2$ in (4) can be excluded as a small parameter, and for the elder
shells
\begin{equation}
i\frac{dy_{n}(\tau )}{d\tau }\simeq 4n^{2}y_{n}(\tau ) \ \ \ (n>>10)\ \ .
\label{12}
\end{equation}
This function is almost independent of the neighbor shells and it
has the solution
\begin{equation}  \label{13}
y_n(\tau )\simeq \mathrm{e}^{-4in^2\tau}y_n(0) \ .
\end{equation} But
$y_n(0)=0$ for any $n\not= 0$, thus, the elder shells do not
participate in the re-distribution of the initial Gaussian
population. Thus, the simple parametric control with the fixed $R$
is not enough to split the beam efficiently. Another scheme of
time-dependent $R$ (corresponding to the most general open-loop
control) should be applied.

\section{ Numerical simulation results for open-loop control with harmonical modulation}
Now let us consider the two level atom in a far detuned standing
wave with an intensity, which is modulated in time harmonically as
$I=I_0(1+\varepsilon \mathrm{cos}(\Delta t))^2\mathrm{cos}^2(kx)$,
where $\varepsilon$ is the amplitude and $\Delta$ is the frequency
of the modulation.\\
We assume also that an initial wave function $\Psi _1(p,\tau =0)$ has
Gaussian profile with the width $\delta p$:
\begin{equation}  \label{14}
\Psi _1(p,\tau =0) = \frac{1}{\sqrt{2\pi}}\mathrm{exp}\left[
-\frac{p^2}{ (\delta p)^2} \right]. \
\end{equation}
We remind that now we use the dimensionless time $\tau =\omega _Rt$.
Fig.2 shows the numerical solution of an equation for amplitude of
the probability of ground state $|\Psi _{1}(p,\tau )|^{2}$ in
momentum representation for the cases unmodulated and modulated
standing wave. We assume that initial wave packet has the width
equals $\delta p=0.5\hbar k$ and $\varepsilon =0.8$, $\Delta /\omega
_{R}=29$. As we can see from this picture, the scattering result
strongly depends on the amplitude modulation exiting in this system.
If for an unmodulated case, it is well-known scattering picture
observed (Fig.2), when an initial wave packet is splitted into a
number of momentum components. However, for modulated standing wave
the scattering picture is changing dramatically and two main
momentum components centered on $\pm 40\hbar k$ can be observed
(Fig.3). Such behavior of the momentum components is due to specific
parametric resonance, which occurs in this system by the well
defined amplitude and frequency modulation. We have to point out
that the values of the modulation obtained for the amplitude and
frequency modulation are strongly different from \cite{2}, and we
can interpret such resonances as Bragg resonances of high orders in
modulated standing wave.

\begin{figure}[h]
 \hskip0cm\centerline{\hbox{
 \includegraphics[clip=true, bb=0 1 10cm 7cm]{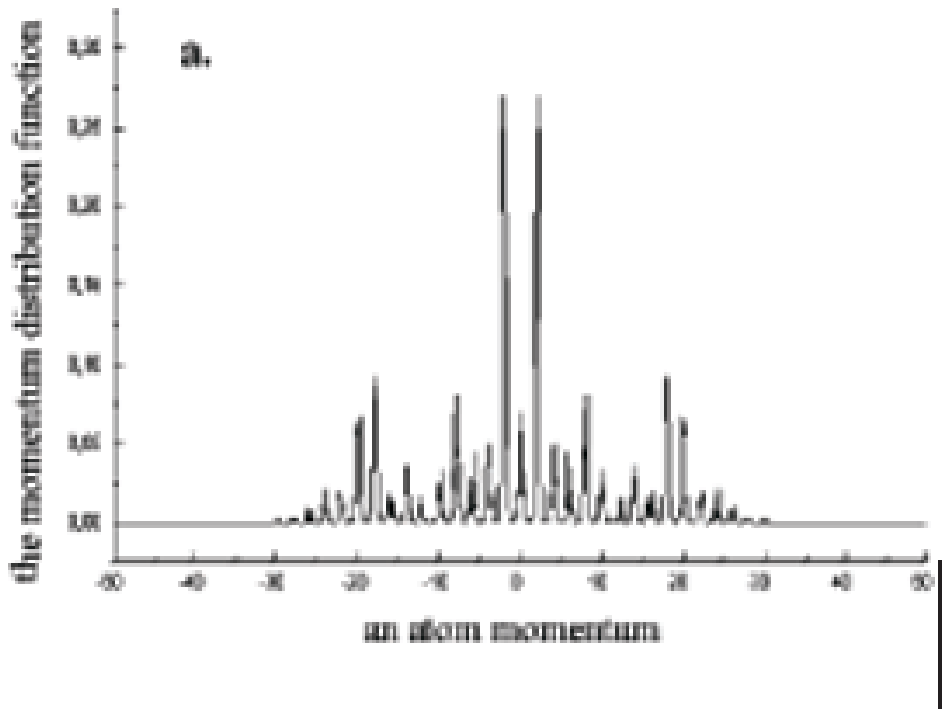}}}
  \caption{The dependence of the distribution function on an atom momentum for an
interaction time $\tau _{\mathrm{int}}=0.567$. The unmodulated standing
wave with the dimensionless Rabi frequency $R_{0}=(320)^{1/2}$. }
\label{fig2}
\end{figure}

\begin{figure}[h]
 \hskip0cm\centerline{\hbox{
 \includegraphics[clip=true, bb=0 1 10cm 10cm]{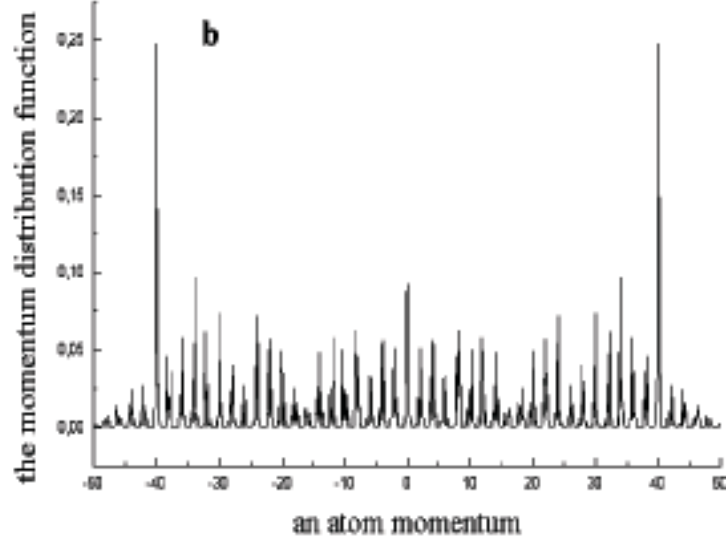}}}
  \caption{The dependence of the distribution function on an atom momentum for an
interaction time $\tau _{\mathrm{int}}=0.567$. The
modulated standing wave with the dimensionless Rabi frequency
$R_{0}=(280)^{1/2}$. }
\label{fig3}
\end{figure}

\newpage
\section{\bf Acknowledgment}
The authors wish to thank Prof. Boris Matisov (St. Petersburg
Polytechnic University) for productive scientific discussions. The
numerical simulation part of this research was financially supported
by the Russian Foundation of Basic Research, Grant 04-02-16175A.

\end{document}